\documentclass[3p,times]{elsarticle}

\usepackage{ecrc}

\volume{00}

\firstpage{1}

\journalname{Nuclear Physics A}

\runauth{F. Gelis}

\jid{npa}

\jnltitlelogo{Nuclear Physics A}

\usepackage{etex}
\usepackage{bbm}
\usepackage[all]{xy}
\usepackage{graphics}
\usepackage{multirow}
\usepackage{array} 

\usepackage{hyperref} 
\usepackage{amsmath}
\usepackage{amsfonts,amsbsy}
\usepackage{amssymb}
\usepackage{bm}

\biboptions{square,comma,numbers,sort&compress}

\usepackage[figuresright]{rotating}

\def\empile#1\over#2{\mathrel{\mathop{\kern 0pt#1}\limits_{#2}}}
\def\bs{\boldsymbol}

\def\TODO#1{}

\def\p{{\boldsymbol p}}
\def\q{{\boldsymbol q}}

\def\k{{\boldsymbol k}}

\def\x{{\boldsymbol x}}

\def\r{{\boldsymbol r}}

\newcommand{\slL}{\raise.15ex\hbox{$/$}\kern-.53em\hbox{$L$}}
\newcommand{\slP}{\raise.15ex\hbox{$/$}\kern-.53em\hbox{$P$}}
\newcommand{\slD}{\raise.15ex\hbox{$/$}\kern-.53em\hbox{$D$}}
\newcommand{\slp}{\raise.1ex\hbox{$/$}\kern-.63em\hbox{$p$}}
\newcommand{\slq}{\raise.1ex\hbox{$/$}\kern-.53em\hbox{$q$}}
\newcommand{\slv}{\raise.1ex\hbox{$/$}\kern-.63em\hbox{$v$}}
\newcommand{\slR}{\raise.15ex\hbox{$/$}\kern-.53em\hbox{$R$}}
\newcommand{\slQ}{\raise.15ex\hbox{$/$}\kern-.53em\hbox{$Q$}}
\newcommand{\slK}{\raise.15ex\hbox{$/$}\kern-.53em\hbox{$K$}}
\newcommand{\slk}{\raise.15ex\hbox{$/$}\kern-.53em\hbox{$k$}}
\newcommand{\slSigma}{\raise.15ex\hbox{$/$}\kern-.53em\hbox{$\Sigma$}}
\newcommand{\slcalP}{\raise.15ex\hbox{$/$}\kern-.63em\hbox{$\cal P$}}
\newcommand{\slcalA}{\raise.15ex\hbox{$/$}\kern-.63em\hbox{$\cal A$}}
\newcommand{\slA}{\raise.15ex\hbox{$/$}\kern-.73em\hbox{$A$}}
\newcommand{\slbfA}{\raise.15ex\hbox{$/$}\kern-.73em\hbox{${\imb A}$}}
\newcommand{\slpartial}{\raise.15ex\hbox{$/$}\kern-.53em\hbox{$\partial$}}
\newcommand{\sla}{\raise.15ex\hbox{$/$}\kern-.53em\hbox{$a$}}
\newcommand{\slb}{\raise.15ex\hbox{$/$}\kern-.53em\hbox{$b$}}
\newcommand{\slc}{\raise.15ex\hbox{$/$}\kern-.53em\hbox{$c$}}
\newcommand{\slC}{\raise.15ex\hbox{$/$}\kern-.63em\hbox{$C$}}

\def\colora{}
\def\colorb{}
\def\colorc{}
\def\colord{}
\def\colore{}

\begin{document}

\begin{frontmatter}



\dochead{}

\title{Theory @ Hard Probes 2013}


\author{Fran\c cois Gelis}

\address{Institut de Physique Th\'eorique, CEA/Saclay, 91191 Gif sur Yvette cedex, France}

\begin{abstract}
This talk presents an overview  of the theoretical contributions at the Hard Probes 2013 conference, held in Stellenbosch, South Africa, in November 2013.
\end{abstract}

\begin{keyword}
Heavy Ion Collisions, Quark Gluon Plasma, Quantum Chromodynamics



\end{keyword}

\end{frontmatter}


\section{Introduction}
\label{sec:intro}
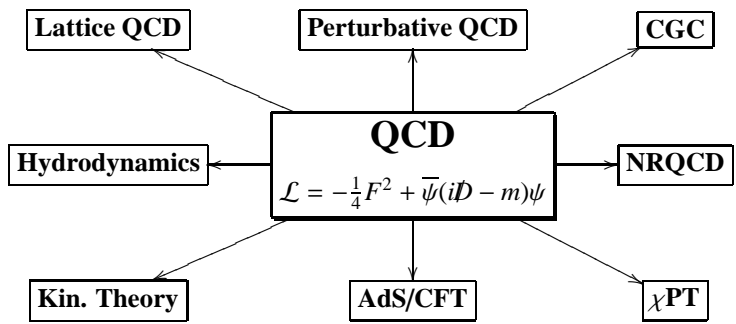
\begin{figure}[htbp]
\xymatrix{
    *+[F-]\txt{{\colorb\bf Lattice QCD}} 
    &
    *+[F-]\txt{{\colorc\bf Perturbative QCD}}
    &
    *+[F-]\txt{{\colord\bf CGC}}
    \\
    *+[F-]\txt{{\colore\bf Hydrodynamics}}
    &
    *+[F-,]{\txt{{\Large\bf QCD}\\{\ }\\${\cal L}=-\frac{1}{4}{\colorb F^2} +{\colora\overline\psi}(i{\colorb\slD}-{\colora m}){\colora\psi}$}}
    \ar[dl]\ar[ul]\ar[dr]\ar[ur]\ar[r]\ar[u]\ar[l]\ar[d]
    &
    *+[F-]\txt{{\colord\bf NRQCD}}
    \\ 
     *+[F-]\txt{{\colore\bf Kin. Theory}} 
    & 
    *+[F-]\txt{{\colora\bf AdS/CFT}}
    &
    *+[F-]\txt{{\bf\colord $\chi$PT}}
}
\caption{\label{fig:overview}Panorama of (some of) the tools used in the theoretical study of heavy ion collisions.}
\end{figure}

Although quantum chromodynamics (QCD) has been recognized as the
microscopic theory of strong interactions for over 40 years, its
application to the study of complex reactions such as heavy ion
collisions remains very challenging. These reactions involve a
complicated mix of various momentum scales, ranging from very hard
scales down to the confinement scale. For this reason, the theory of
heavy ion collisions is fragmented into many different tools,
effective descriptions and models, as illustrated in the figure
\ref{fig:overview}. 
Arguably, the closest to QCD are lattice QCD and perturbation theory,
but both are seriously limited in what questions they can
address. Somewhat further away from QCD, one encounters some effective
theories that approximate the QCD dynamics in a certain regime (high
gluon density, heavy quarks, ...). Calculations based on the AdS/CFT
correspondence allow one to study the strong coupling limit of certain
cousin gauge theories, like the SUSY ${\cal N}=4$ Yang-Mills theory.
Finally, transport descriptions can be used at larger distance scales
where the microscopic details are not so important and the dynamics is
largely controlled by conservation laws.

Another important aspect of heavy ion collisions is that successful
descriptions must include some mundane facts that have no
connection with the aspects of QCD that motivate studying heavy-ion
collisions, such as the geometry of nuclei, how the nucleons are
distributed inside nuclei and how they fluctuate, etc... All these
things matter in the interpretation of heavy ion data, before one can
get at the underlying QCD phenomena.  The number of participants gives
a good idea of the geometry of a given collision. It is not directly
measurable, but in nucleus-nucleus collisions it is well correlated to
some measurable quantities, such as the transverse energy or the
deposit in zero degree calorimeters. A worrisome lesson from HP2013 is
the fact that our handle on the geometry of proton-nucleus collisions
is not very good. In these collisions, there is a rather weak
correlation between the multiplicity and the centrality of the
collision, which makes using pA collisions as a reference for AA
collisions rather problematic. Because of this, understanding pA
collisions has become a serious challenge in itself, quite bad news
if one wishes to use them as a reference.

With this caveat in mind, the general idea of {\sl hard probes} is to
measure objects characterized by some large momentum scale (e.g. jets,
photons, bound states containing heavy quarks), and to infer
properties of the medium in which they propagate by studying how some
of their properties are modified compared to collisions where one does
not expect the formation of a quark-gluon plasma (e.g. proton-nucleus
and proton-proton collisions). The theory contributions to HP2013 can
be divided in roughly four topics~:
\begin{itemize}
\item[{\bf 1.}] {\bf Initial state.} {\it(16 contributions)} This
  encompasses the very early stages of a collision, while the system
  is still dense enough to be characterized by a semi-hard momentum
  scale. This departs a bit from the general concept of hard probes,
  because in these studies the medium itself is treated
  perturbatively.
\item[{\bf 2.}] {\bf Electromagnetic probes.} {\it(7 contributions)}
  Photons and dileptons, even when their momenta are not very large,
  are usually considered as part of the ``hard probes'' because their
  weak (electromagnetic only) interactions give them a mean free path
  much larger than the system size. Their spectrum therefore reflects
  the state of the system at the time of their production.
\item[{\bf 3.}] {\bf Energy loss.} {\it(16 contributions)}
  Modifications of inclusive (identified or not) hadron spectra have
  been observed long ago in heavy ion collisions. More recently, these
  measurements have been extended to jets, which opens up new
  possibilities for studying the medium modifications of radiation.
\item[{\bf 4.}] {\bf Heavy flavors.} {\it(11 contributions)} Rather
  unexpectedly, heavy quarks tend to lose a significant amount of
  energy in a dense medium, like their lighter siblings. Moreover,
  compact bound states made of heavy quarks can probe the local
  temperature of the medium.
\end{itemize}
The number of contributions in each of these subjects can be viewed as
a vague indication of where the field is going, and what topics are
considered to be in need of more work, or on the contrary are getting
closer to a satisfactory state of understanding.

\section{Initial state}
\label{sec:init}
The first step in studying hard probes is to determine their initial
yield. In the case of jets and heavy flavors (for quark masses that
are large compared to the saturation momentum), this can be done in
the DGLAP framework, using nuclear parton distribution functions
(NPDFs). In addition to having some shadowing (i.e. they are not a
simple superposition of nucleon PDFs), these distributions can also be
made transverse position dependent in order to account for the fact
that the nuclear modification may depend on the local density of
nucleons ({\sc I. Helenius}, \cite{Helenius:2014cva}).

When the typical momentum scale of the process under consideration
is comparable to the saturation momentum, recombination
corrections become important and cannot be included in the DGLAP
framework. However, these effects can somehow be mimicked by
introducing a form of ``final state saturation'' via an infrared
cutoff. This is the essence of the EKRT model, that has now been
extended to include NLO partonic cross-sections ({\sc R. Paatelainen},
\cite{Paatelainen:2012at}).

At high transverse parton density, double parton scatterings can occur
in a single collision. Ignoring quantum coherence effects, this can be
modeled in terms of single parton scattering cross-sections
\cite{d'Enterria:2014bga}.  However, the Color Glass Condensate is a
more rigorous and systematic way to treat the physics of multi-parton
interactions in QCD. It is based on the separation of the partons into
``fast'' partons that are described as light-cone color currents
$J_{1,2}$ (one for each colliding projectile) and ``slow'' partons
that are described by the usual quantum fields, glued together by an
eikonal coupling~: $ {\cal L} = -\frac{1}{4}{\colorc F_{\mu\nu}
  F^{\mu\nu}} + {\colorc A_\mu}\cdot ({\colorb J^\mu_1}+{\colora
  J^\mu_2})\; .  $ The currents $J_{1,2}$ reflect the instantaneous
transverse positions and colors of the fast partons at the time of the
collision. This can only be described probabilistically, with
distributions $W[J_{1,2}]$ that depend solely on the nature of the
projectile, and on the scale $\Lambda$ that separates the fast and
slow partons. Like with conventional PDFs, the distributions
$W[J_{1,2}]$ are non-perturbative, but their scale dependence is
driven by a perturbative equation, the JIMWLK equation in the case of
the CGC.  Since the JIMWLK equation is a functional equation whose
numerical solution is very time consuming, it is often replaced by a
mean-field approximation, the BK equation, in phenomenological
applications. The BK equation can be viewed as an approximation scheme
that closes the equation for the dipole amplitude (that would
otherwise depend on a 4-point function). More general truncation
schemes have been proposed, such as the Gaussian truncation ({\sc
  A. Ramnath}, \cite{Ramnath}, {\sc G. Jackson}, \cite{Jackson}).

Due to the simplicity of their implementation, models based on the BK
equation have been used to describe proton-nucleus collisions, where
it is more legitimate to neglect final state effects. This can be done
using $k_T$-factorization and BK evolution for the two projectiles
({\sc H. Mantysaari}, \cite{Lappi:2014pua}) or in a hybrid formulation
where the BK evolution is applied only to the nucleus, while the
proton is described with collinear factorization ({\sc B.-W. Xiao},
\cite{Xiao:2014pla}).

The BK equation had also the advantage of being well known at
next-to-leading log accuracy. However, it is now becoming possible to
include running coupling effects in the Langevin formulation of the
JIMWLK equation ({\sc T. Lappi}, \cite{Lappi:2014wya}). Moreover, a
full NLL JIMWLK equation has recently appeared on the horizon ({\sc
  S. Caron-Huot}, \cite{Caron-Huot:2013fea} and {\sc M. Lublinsky,
  A. Kovner, Y. Mulian}, \cite{Kovner:2013ona}). Interestingly, simple improvements
in the way the kinematics is treated in leading log evolution
equations can account for a significant part of the NLL corrections
({\sc G. Beuf}, \cite{Beuf:2014uia}).

Another recent improvement to the JIMWLK equation is the possibility
to tag some of the gluons that are radiated in the course of the
evolution ({\sc E. Iancu}, \cite{Iancu:2013uva}), in order to compute
multi-gluon correlations (this is necessary for instance when studying
ridge-like correlations\footnote{We were reminded by {\sc A. Kovner}
  \cite{Kovner:2012jm} of some mechanisms that can produce the
  azimuthal collimation of the ridge without invoking any flow.} in
proton-nucleus collisions).  The practicality of such a generalized
evolution equation is somewhat limited at the moment, since the
required computational resources would scale as $S_\perp^{\#\ {\rm
    gluons}}$, where $S_\perp$ is the number of lattice sites in the
discretization of the transverse plane.

In nucleus-nucleus collisions, the expansion of the bulk of the matter
is usually described by relativistic hydrodynamics, and models such as
the CGC are used only to describe the pre-hydrodynamical stages of the
collision, and to provide some initial data (energy density, pressure,
flow velocity and stress tensor at some initial time). Besides the
CGC, there are many models exist to compute these quantities with
various degrees of sophistication ({\sc S. Jeon}, \cite{Jeon})~:
Optical Glauber, MC Glauber, AMPT, NEXUS, MC-KLN, MC-rcBK, IP-GLasma
(ordered here by the amount of QCD they contain). These models differ
in the spatial scales at which they have fluctuations. In principle,
by studying the event-by-event mapping from the initial shape
harmonics $\epsilon_n$ to the final flow harmonics $v_n$, one could
constrain the parameters that define these models, and at the same
time determine transport coefficients such as the shear viscosity.

At Leading Order, the pressure tensor predicted by the CGC remains
very anisotropic at all times. Since this is hard to accommodate within
hydrodynamics, the matching to hydrodynamics usually uses only the CGC
energy density and assumes an equilibrium equation of state to guess
the corresponding pressure. However, it seems that resumming some
higher order corrections in the CGC calculation has an important
effect on the longitudinal pressure ({\sc T. Epelbaum},
\cite{Gelis:2013rba}). In addition, by a time comparable to the
inverse saturation momentum, a non-zero flow velocity has developed
({\sc R. Fries}, \cite{Chen:2013ksa}), made of a term proportional
to the gradient of the energy density, and a term specific to
Yang-Mills fields that involves field commutators. This initial flow
should certainly be part of the hydrodynamical initial conditions, if
one wants the final results to be as insensitive as possible to the
time at which it is initialized.

Somewhat more on the exploratory side, there have been some attempts to
mimic a nucleus-nucleus collision by colliding two shockwaves in the
strong coupling regime of the SUSY $N=4$ Yang-Mills theory, using the
AdS/CFT correspondence. The energy of the collision can be controlled
by tuning the thickness of the shockwaves. By varying this parameter,
a transition from a situation close to Landau hydrodynamics (at low
energy) to  Bjorken-like hydrodynamics (at high
energy) is observed ({\sc J. Casalderrey-Solana}, \cite{Casalderrey-Solana:2013aba}).

\section{Electromagnetic probes}
\label{sec:em}
The production rate of photons and low mass dileptons by a hot plasma
increases like $T^4$, which means that these probes are very sensitive
to the temperature of the medium at the earliest stages. Moreover,
their mean free path is much larger than the size of the system and
therefore they can escape without any modification of their spectra.
The only drawback of photons as probes of the quark-gluon plasma is
that there are many other sources of photons: prompt photons produced
at the impact of the two nuclei, before the formation of the QGP,
photons produced by the hot hadronic gas that forms after the
confinement transition, and a copious background of photons from the
decay of unstable hadrons in the final state, e.g. $\pi^0$'s.

The calculation of the photon and dilepton production rates in a
thermal medium can be done by calculating the relevant squared matrix
elements, integrated over the phase-space of the unobserved particles
with the appropriate statistical factors. Alternatively, the
bookkeeping can be made more systematic by relating the photon
production rate to the imaginary part of the photon polarization
tensor, that can be calculated in QCD at finite temperature by using
cutting rules.

The lowest order processes ($q\bar{q}\to\gamma^*$ for dileptons,
$q\bar{q}\to \gamma g$ and $qg\to \gamma q$ for real photons) have
first been calculated long ago. In the last two processes, the quark
exchanged in the t-channel can become soft, which makes the rate
formally infinite. This is cured by thermally generated quark masses
(resummed via the Hard Thermal Loops effective theory), which turns
the infinity into a logarithm.

However, it was realized shortly afterwards that the higher order
bremsstrahlung graphs also have the same order of magnitude, due to a
strong collinear singularity regularized by a quark thermal
mass. Physically, this divergence arises when the photon formation
time becomes larger than the quark mean free path, and the cure
involves resumming multiple scattering corrections, in order to
capture the Landau-Pomeranchuk-Migdal interference effect. The full
leading order photon and dilepton rates have been obtained in
\cite{Arnold:2001ms} and \cite{Aurenche:2002wq}.

Major improvements over these old results have been performed recently
and were presented at HP2013. Firstly, the production rate of real
photons has now been extended to NLO in the strong coupling constant
({\sc J. Ghiglieri}, \cite{Ghiglieri:2013gia}). This calculation is
interesting at the technical level since several new tricks were
invented on this occasion, and the result itself is quite remarkable
because it is very close to the old LO result, due to a partial (and
purely accidental) cancellation between two corrections of opposite
signs.

The thermal dilepton rate has also been calculated at NLO ({\sc
  M. Laine}, \cite{Laine:2013vma}) for pair invariant masses that are
large enough so that the resummation of the multiple scattering
corrections is not necessary. Here it was observed that the NLO
corrections become very large compared to the LO contribution at small
invariant masses, possibly due to new production channels opening up
at NLO. Interestingly, this calculation shows a fairly good agreement
with correlators evaluated on the lattice at zero spatial momentum.

Let us now digress on the collision kernel ${\cal C}(q_\perp)$ that enters in
the calculation of photon rates and low mass dilepton rates. In short,
${\cal C}(q_\perp)$ is the squared scattering amplitude of the probe off a
particle of the medium, integrated over the phase-space of the
scattering center with the appropriate statistical distribution. The
t-channel propagator in the amplitude must also be dressed for medium
effects such as Debye screening. This object also appears in the
discussion of the transverse momentum broadening undergone by a fast
parton moving through a thermal medium, since ${ \Delta
  k_\perp^2 \mbox{\ per collision} =
  \int\frac{d^2\q_\perp}{(2\pi)^2}\,q_\perp^2\;{\cal C}(\q_\perp)}$. 
Perturbatively, this collision kernel can be expressed in terms of
in-medium gluon spectral functions~:
\begin{equation*}
{\colorb{\cal C}(\q_\perp)}=
\int
{{dq_0 dq_z}\over{(2\pi)^2}}\,2\pi\delta(q_0-q_z)
\;{\colora\widehat{v}_\mu  \widehat{v}_\nu}\;\Big({\colorb \rho_{_L}^{\mu\nu}(Q)}+{\colord \rho_{_T}^{\mu\nu}(Q)}\Big)\; .
\end{equation*}
From this expression and the Hard Thermal Loop results for the gluon
spectral functions, one obtains the following LO expression~\cite{Aurenche:2002pd}, ${\cal
  C}(q_\perp) = {m_{_D}^2}/({q_\perp^2}({q_\perp^2+{\colorb
    m_{_D}^2}}))$ (where $m_{_D}$ is the Debye mass). In order to go
beyond this result, one may use the following non-perturbative
definition
\begin{equation*}
e^{-\ell\,{\widetilde{\cal C}(r_\perp)}}
=
\frac{1}{N_c}\,{\rm Tr}\,(U(0_\perp)U^\dagger(\r_\perp))
\end{equation*}
that relates the Fourier transform $\widetilde{\cal C}(r_\perp)$ of the collision kernel to a
correlator of Wilson lines along the light-cone. The NLO collision
kernel has been calculated in \cite{CaronHuot:2008ni}. Recently, this
formula has been used in conjunction with the EQCD effective theory,
obtained by dimensional reduction after integrating out all the modes
with a non-zero Matsubara frequency, ({\sc M. Panero},
\cite{Panero:2013pla}) in order to calculate ${\cal C}(r_\perp)$ on the
lattice. This computation leads to an estimated value $\hat{q}\approx
6~$GeV${}^2/$fm for the jet quenching parameter at typical RHIC
temperatures. Interestingly, if scaled appropriately by the Debye
mass, this lattice result agrees quite well with the NLO perturbative
result at small distance $r_\perp$.

In the context of heavy quarks energy loss, it was proposed by {\sc
  M. Djordjevic} \cite{Djordjevic:2012qp} to modify the LO collision kernel as follows
\begin{equation*} 
 \frac{1}{q_\perp^2}-\frac{1}{q_\perp^2+{\colorb m_{_D}^2}}
\quad\to\quad
\frac{1}{q_\perp^2+{\colord m_{_M}^2}}-\frac{1}{q_\perp^2+{\colorb m_{_D}^2}}
\end{equation*}
in order to include phenomenologically a magnetic screening mass
$m_{_M}$ (lacking at LO). Although this modification seems to be
helpful with the phenomenology of energy loss, one should keep in mind
the fact that magnetic screening is intrinsically non-perturbative in
QCD, and cannot be encompassed completely by this mere change in
otherwise leading order formulas.

The local production photon and dilepton production rates are then
folded into a hydrodynamical evolution in order to perform the space
and time integrations that leads to the produced spectra. Basically,
one just needs to boost the rates in the rest frame of each fluid cell
by the flow velocity of that cell. When one fits the measured spectra
by an exponential of the form $\exp(-p_\perp/T)$, it is important to
keep in mind that the slope parameter is not the initial temperature
of the quark-gluon plasma ({\sc U. Heinz}, \cite{Heinz:2014uga}), due
to the integration over the whole space-time history of the system.

When using viscous hydrodynamics in this procedure, one is really
considering an out-of-equilibrium system: a non-zero viscous stress
tensor $\pi^{\mu\nu}$ implies that the distributions $f(p)$ depart
from the equilibrium ones. Consequently, the photon and dilepton rates
themselves should be corrected to account for these modified quark and
gluon distributions. This has been done for the $2\to 2$ production
processes ({\sc C. Shen}, \cite{Heinz:2014uga}), and should be done also for
the multiple scattering corrections to bremsstrahlung.  It also
seems that the photon $v_2$ exhibits some sensitivity to the
parameters of the early flow ({\sc G. Vujanovic},
\cite{Vujanovic:2014xva}).

When one compares the photon spectra generated by these models to the
data from nucleus-nucleus collisions, it appears that the predicted
yield is a bit too small. But the largest discrepancy is in the $v_2$
of the photons, whose predicted value is too small by a factor $\sim
4$. The conjunction of these two facts suggests that the present
models may underpredict photon production in the hadronic phase, where
the flow is already developed (for a step in this direction, see {\sc
  W. Cassing}, \cite{Linnyk:2013wma}).

\section{Energy loss and jet quenching}
\label{sec:Eloss}
The modifications of the propagation and fragmentation of a parton
when it propagates through a dense medium is a natural way to probe
the properties of this medium. These medium modifications depend on
a wide range of scales, going from the initial energy of the probe to
the Debye screening scale in the medium, which makes it rather
complicated to describe in detail. This complexity has lead to several
popular approaches (BDMPS, GLV, AMY,...), where the emphasis is put on
different aspects of the problem. A recent trend has been to go beyond
these original models, especially the BDMPS one, in order to include
physics that was not in there originally ({\sc C. Salgado},
\cite{Salgado}, {\sc E. Iancu}, \cite{Iancu:2014aza}).

The common ground of all these models is soft and collinear gluon
radiation in QCD,
${dP}\sim \alpha_s\; {\colorb ({dx}/{x})}{\colord ({d^2\k_\perp}/{k_\perp^2})}\; ,
$
which is enhanced by the momentum kicks that the probe receives while
propagating through a medium. For independent scatterings, the
transverse momentum of the probe increases as $k_\perp^2 \sim
\hat{q}\, t_f$ (in this discussion, $\hat{q}$ is the only parameter
characterizing the medium), while the gluon formation time is
$t_f\sim{\omega}/{k_\perp^2}$. Therefore, $t_f\sim
\sqrt{{\omega}/{\hat{q}}}$.  For induced radiation to happen, $t_f$
should be less than the medium size, which puts an upper limit
${\colord \omega_c= \widehat{q}\,L^2}$ on the energy of the radiated
gluons. The angular aperture of these emissions is given by
$\theta\sim {k_\perp}/{\omega}\sim
\big({\hat{q}}/{\omega^3}\big)^{1/4}$, which means that the hardest
gluons are emitted at small angles. The mean energy lost in a single
emission is $\big<\omega\big>\sim \alpha_s\omega_c \sim \alpha_s
\widehat{q}\,L^2$, i.e. it is dominated by $\omega$ close to the upper
limit $\omega_c$ and it stays in a small cone around the trajectory of
the probe.  Therefore, this explains the energy loss of single
hadrons, but it does not explain why jets are quenched.

The suppression of jets requires that energy be radiated at large
angles (outside of the jet cone). This can happen for soft emissions,
but many of them will be necessary to achieve a significant
suppression.  The color fields of the medium tend to randomize the
colors of the emitters, which destroys the vacuum coherent antenna
effects for out-of-cone emissions.  Because of this, large angle
emissions are not suppressed, and the energy radiated out-of-cone is
rapidly degraded all the way down to the medium temperature
scale. In-cone, the medium does not resolve the individual jet
constituents, and the jet fragmentation proceeds as in vacuum.

Another effect is the correction to the parameter $\hat{q}$ due to the
fact that the gluon emissions contribute to the accumulated transverse
momentum ({\sc Y. Mehtar-Tani}, \cite{Blaizot:2014bha}). This effect
is suppressed by $\alpha_s$ but enhanced by logarithms, and if
resummed it introduces an anomalous $L$ dependence in $\hat{q}$~:
$\big<\omega\big>\sim \alpha_s \widehat{q}_0\,{\colorb L^2}
\big({\widehat{q}_0 {\colorb
    L}}/{m_{_D}^2}\big)^{\colorc\sqrt{4\alpha_s
    N_c/\pi}}$. Interestingly, this result is between the BDMPS result
$L^2$ and the strong coupling result $L^3$. Several technical
improvements were also presented, such as the complete single gluon
radiation off a quark, including the in- and out-of-medium emission
and their interference ({\sc L. Apolinario}, \cite{Apolinario:2014xla}) and a
study of the effect of the mass of heavy quarks on the destruction of
the antenna coherence ({\sc M. Calvo}, \cite{Calvo:2014cba}).

Phenomenological models of energy loss that accommodate a range of
behavior from elastic collisions only to the strong coupling result,
embedded in viscous hydrodynamics, tend to prefer an energy loss that
varies with a power law close to $L^2$, while stronger $L$ dependences
are disfavored ({\sc B. Betz}, \cite{Betz:2014ooa}). This may not
exclude totally strong coupling scenarios, since it appears that
shooting string holographic models of jet quenching can also produce a
range of powers of $L$ ({\sc A. Ficnar},
\cite{Ficnar:2014zba}). Another way to circumvent the shortcomings of
plain AdS/CFT is a hybrid strong/weak coupled approach to jet
quenching ({\sc D. Pablos}).

The last step in more realistic studies is to embed a given
microscopic model of energy loss in a model for the expansion of the
fireball. For instance, the $v_2$ is quite sensitive to the details of
the latter ({\sc D. Molnar}, \cite{Molnar:2012fn}). How one chooses
the triggers is also important, as this can bias the event sample in
several ways (chemical composition, kinematics, location of the
initial hard scattering,..). This may be used to select specially
interesting situations, but it also makes the comparison to data more
complicated ({\sc T. Renk}, \cite{Renk:2014pba}). A number of real
world implementations were also presented at HP2013, that can be
roughly categorized into hydro-based models ({\sc J. Xu},
\cite{Xu:2014wua}, {\sc Y. Tachibana}, \cite{Tachibana:2014vra}, {\sc
  R. Andrade}, \cite{Andrade:2014swa}) or cascade-like models ({\sc
  J. Tan}, \cite{Tan}, {\sc Y. Zhu}, \cite{Wang:2014hla}, {\sc
  K. Tywoniuk}, \cite{Mehtar-Tani:2014yea}, {\sc K. Zapp}, \cite{Zapp:2013zya}).

\section{Heavy flavors}
\label{sec:HF}
An important subtopic in the field of heavy flavors is that of energy
loss. Until recently, there was an apparent contradiction between the
observation that heavy quarks seem to loose almost as much energy as
lighter quarks, and the theoretical expectation that the dead-cone
effect forbids radiation at angles $\theta<M/E$. This observation
triggered a lot of interest for strong coupling models based on
AdS/CFT, and a revived interest in collisional energy loss.

It now appears that, by carefully taking the partonic energy loss
(both radiative and collisional, and by including running coupling as
well as a phenomenological form of magnetic screening), the
fragmentation functions and the subsequent decay of the D mesons, one
can now obtain a reasonably good agreement with the suppression data
for inclusive hadrons and for D mesons ({\sc M. Djordjevic},
\cite{Djordjevic:2013pba}).  This improvement seems due to several
smaller changes in various places, rather than having a unique cause.

The energy loss of heavy quarks can also be studied in cascade models
({\sc C. Greiner, J. Uphoff}, \cite{Uphoff:2013soa}). In order to
account for the radiative processes, it is necessary to include $2\to
3$ and $3\to 2$ processes as well as $2\to 2$ processes, and the LPM
interference is mimicked by vetoing rescatterings during the formation
time of the emitted gluons. For inclusive hadrons, this leads to a
reasonable $R_{_{AA}}$, while for heavy quarks some tension persists
between $R_{_{AA}}$ and $v_2$.

Hybrid models mixing transport ideas and an underlying hydrodynamical
background were also presented at HP2013 ({\sc M. Nahrgang,
  P.B. Gossiaux }, \cite{Nahrgang:2013pka}). In this particular model,
EPOS was used for the description of the bulk, and the initial heavy
quark distribution was taken from FONLL. A similar idea was presented
by {\sc M. Nardi}, \cite{Alberico:2013bza}. Another, somewhat related,
approach is to modify the Langevin equation that describes the
momentum diffusion of heavy quarks by adding a force term that mimics
the radiative energy loss ({\sc S. Cao}, \cite{Cao:2014pka}).

In the strong coupling limit, the drag force that a medium exerts on a
heavy quark has been known for a long time,
${d\vec\p}/{dt}=({\sqrt{\lambda}}/{2\pi})(\pi T)^2
{{\bs\beta}}/{\sqrt{1-\beta^2}}$, in the case of a static and
isotropic quark-gluon plasma. A more sophisticated study in the case
of a medium that has an anisotropic energy-momentum tensor, as would
be the case for a longitudinally expanding quark-gluon plasma,
indicates that the magnitude of the drag force is not dramatically
modified, with only relatively small changes due to the anisotropy
({\sc M. Lekaveckas }, \cite{Lekaveckas:2013lha}).

The second subtopic related to heavy quarks is that of quarkonia
dissociation in a hot and dense medium. There is now a clear signal at
the LHC of the suppression of the excited states of the $\Upsilon$,
which is very suggestive of a sequential dissociation pattern, where
the ground state melts at a higher temperature than the excited ones.

A lot of activity in lattice QCD is devoted to the study of quantities
relevant to the physics of heavy quarks in the quark-gluon plasma. A
notoriously difficult problem is the calculation of a spectral
function $\sigma$. On the lattice, the closest object one can compute
is an Euclidean correlator $G_{_E}$, related to $\sigma$ by
\begin{equation*}
{\colorb G_{_E}(\tau,\p)}
=\int\limits_0^\infty \!\! d\omega\, {\colora\sigma(\omega,\p)}
\,
\frac{\cosh(\omega(\tau-\beta/2))}{\sinh(\omega\beta/2)}\; .
\end{equation*}
The inversion of this integral relationship is an ill-posed problem,
unless one introduces some extra informations or assumptions about the
spectral function, which can be done with the maximum entropy
method. An additional difficulty is that the Euclidean correlator
$G_{_E}$ is not very sensitive to changes of the spectral function
({\sc H.-T. Ding},\cite{Ding}). This method can be used to extract the
spectral function in various channels with heavy quarks, in order to
directly visualize the disappearance of the bound state peaks as the
temperature increases. In the case of the $J/\Psi$, it appears that the
peak is almost gone at $T=1.45 T_c$. This method can also be used to
compute transport coefficients such as the diffusion constant of heavy
quarks. The same approach with light quarks can give access to the
electrical conductivity of the quark-gluon plasma, and to the dilepton
production rate.

The previous method is more and more difficult to apply as the
temperature increases, because the extent of the imaginary time axis
shrinks as $1/T$. This is especially problematic for extracting
transport coefficients, that are obtained from the zero momentum limit
of the spectral function. An alternative approach is to consider an
Euclidean correlator in the spatial $z$ direction, whose extent is not
limited by the temperature,
\begin{equation*}
G(z)\equiv
\int_0^\beta d\tau\;\int dx dy\;
\left<J(\tau,\x)J(0,{\bs 0})\right>
=
\int dz \,e^{ip_z z}\,
\int\limits_0^\infty \!\! d\omega\,
\frac{{\colorb\sigma(\omega,(0,0,p_z))}}{\omega}\; .
\end{equation*}
This alternate method is more sensitive to changes of the spectral
function ({\sc A. Mocsy}, \cite{Mocsy:2013syh}), and high-temperature
modifications become much more obvious.

One can also address the fate of quarkonia made of heavy quarks in
effective theories, such as pNRQCD,
  \begin{equation*}
    {\cal L}=\underbrace{-\frac{1}{4}F_{\mu\nu}F^{\mu\nu}+i\overline{q}\slD\, q}_{\scriptsize\mbox{light quarks and gluons}}
+
{\colorb S^\dagger}\left[i\partial_t -\frac{(i{\bs\nabla})^2}{M}-{\colora V_s(r)}\right] {\colorb S}
+\cdots
  \end{equation*}
where $S$ is an effective field for singlet $Q\overline{Q}$
states. The potential $V_s$ is generally complex, with real and
imaginary parts respectively related to the shift of the peak and the
broadening of the spectral function. In this effective description at
lowest order, the field $S$ obeys a non-relativistic Schrodinger
equation with the potential $V_s$. In the past, the singlet free
energy was used in this equation, but it has no imaginary part. A more
rigorous derivation is from the spectral function ({\sc P. Petreczky
}, \cite{Bazavov:2014kva}). First results indicate that the real part
of the potential is between the zero temperature potential and the
singlet free energy, with a non negligible imaginary part. When this
potential is used in the Schrodinger equation, one finds that all the
charmonium states disappear for $T>240~$MeV. In the case of the
Upsilon states, the $\Upsilon(2s)$ and $\Upsilon(3s)$ melt around
$T\sim 250~$MeV, while the ground state $\Upsilon(1s)$ survives until
$T\sim 450~$MeV.
\vglue 2mm
\noindent{\bf Summary}\ \ 
This 6th edition of the Hard Probes conference has seen many new
results, both in hardcore theory and in the modeling of
nucleus-nucleus collisions. Some areas of the theory of heavy ion
collisions are close to reaching a state where the corresponding data
becomes understandable from a QCD perpective. This conference also
shed some light on the fact that proton-nucleus collisions are not as
simple as one may have thought originally, with many issues related to
the centrality determination, making them more difficult to use
reliably as a control experiment.
\vglue 2mm
\noindent{\bf Acknowledgements}\ \ It is a pleasure to thank the
organizers of the HP2013 conference for the very enjoyable
atmosphere of this meeting. My work is supported by the Agence
Nationale de la Recherche project 11-BS04-015-01.






\end{document}